\documentstyle[12pt]{article}

\newcommand{\cl}{\centerline}
\renewcommand{\theequation}{\arabic{equation}}
\newcommand\beq{\begin{equation}}
\newcommand\eeq{\end{equation}}
\newcommand\bea{\begin{eqnarray}}
\newcommand\eea{\end{eqnarray}}

\begin{document}

\begin{titlepage}
\setlength{\textwidth}{5.0in}
\setlength{\textheight}{7.5in}
\setlength{\parskip}{0.0in}
\setlength{\baselineskip}{18.2pt}
\hfill SOGANG-HEP 257/99
\begin{center}
{\Large{{\bf Improved BFT quantization of}}}\par
\cl{\Large{{\bf $O(3)$ nonlinear sigma model}}}\par
\vskip 0.5cm
\begin{center}
{Soon-Tae Hong$^a$, Won Tae Kim$^b$, and Young-Jai Park$^{c}$}\par
\end{center}
\vskip 0.4cm
\begin{center}
{Department of Physics and Basic Science Research Institute,}\par
{Sogang University, C.P.O. Box 1142, Seoul 100-611, Korea}\par
\end{center}
\vskip 0.5cm
\cl{\today}
\vfill
\begin{center}
{\bf ABSTRACT}
\end{center}
\begin{quotation}
We newly apply the improved Batalin-Fradkin-Tyutin(BFT) Hamiltonian method
to the O(3) nonlinear sigma model, and directly obtain the compact form of 
nontrivial first class Hamiltonian by introducing the BFT physical fields.
Furthermore, following the BFV formalism, we derive the BRST invariant gauge 
fixed Lagrangian through the standard path-integral procedure.
Finally, by introducing collective coordinates, we also study a semi-classical 
quantization of soliton background to conclude that the spectrum of zero modes 
are unchanged through the BFT procedure.

---------------------------------------------------------------------\\
\noindent
$^a$sthong@physics3.sogang.ac.kr\\
\noindent
$^b$wtkim@ccs.sogang.ac.kr\\
\noindent
$^c$yjpark@ccs.sogang.ac.kr

\vskip 0.5cm
\noindent
\end{quotation}
\end{center}
\end{titlepage}

\newpage

\section{Introduction}

The O(3) nonlinear sigma model has interesting physical content in its
phenomenological aspects. For instance, in the Euclidean space it describes
classical (anti)ferromagnetic spin systems at their critical points\cite
{eucl}, while in the Minkowski one it delineates the long wavelength limit
of quantum antiferromagnets\cite{min}. Also the model exhibits solitons,
Hopf instantons and novel spin and statistics in 2+1 space-time dimensions
with inclusion of the Chern-Simons term\cite{cs, bo}.

On the other hand, it has been well known that in order to quantize physical
systems with constraints, Dirac method\cite{di} has been conventionally
used. The Poisson brackets in the second-class constraint system can be
converted into the Dirac brackets for the self-consistency. However, Dirac
brackets are generically field-dependent, nonlocal, and have ordering
problems between field operators. These are unfavorable in finding
canonically conjugate pairs. However, if the first class constraint system is
realized, the usual Poisson bracket corresponding to the quantum commutator
can be used without introducing the Dirac brackets. Quantizations in this 
direction have been well appreciated in a gauge invariant fashion by using the 
Becci-Rouet-Stora-Tyutin(BRST) symmetry \cite{brst}.

To overcome the above problems, Batalin, Fradkin, and Tyutin (BFT) developed a
method \cite{BFT} which converts the second class constraints into first
class ones by introducing auxiliary fields. Recently, this BFT formalism has
been applied to several interesting models \cite{BFT1,kpr}. In particular,
the relation between the Dirac scheme and BFT one, which had been obscure
and unsettled, was explicitly clarified in the framework of the SU(2) Skyrmion 
model\cite{sk2}. Also the BFT Hamiltonian method was applied\cite{su2bft} to the
SU(2) Skyrmion model to directly obtain the compact form of the first class 
Hamiltonian, via the construction of the BFT physical fields.  On the other 
hand, this BFT approach was applied to O(3) nonlinear sigma model by several 
authors, but they could not obtain the desired compact form of the first class 
Hamiltonian\cite{bgb} and therefore could not carry out further discussions.

The motivation of this paper is to systematically apply the improved BFT\cite
{kpr} scheme, and the Batalin, Fradkin and Vilkovisky (BFV)\cite{bfv,fik,biz} 
and BRST methods to the O(3) nonlinear sigma model\cite{bo} as a nontrivial 
example of topological system. In section 2, we convert second class 
constraints into first class ones according to the BFT method.  In section 3, 
we construct the first class BFT physical fields and directly derive the 
compact form of the first class Hamiltonian in terms of these fields without 
studying infinitely iterated standard procedure, which has been carried out 
for the SU(2) Skyrmion case\cite{sk}. Then, we investigate some properties of 
Poisson brackets of these BFT physical fields, which yield the Dirac brackets 
in the limit of vanishing auxiliary fields in section 4. We construct the BRST 
invariant gauge fixed Lagrangian in the BFV scheme through the standard 
path-integral procedure in section 5.  We finally study a semi-classical 
quantization of soliton by introducing collective coordinates in section 6.


\section{First class constraints}
\setcounter{equation}{0}
\renewcommand{\theequation}{\arabic{section}.\arabic{equation}}

In this section, we perform the standard BFT procedure for the O(3)
nonlinear sigma model which is a second class constraint system. In the $O(3)
$ nonlinear sigma model, the starting Lagrangian is of the form 
\begin{equation}
L_{0}=\int {\rm d}^{2}x \left[ \frac{1}{2f}(\partial_{\mu}n^{a})(\partial^{%
\mu}n^{a})\right]
\end{equation}
where $n^{a}$ ($a$=1,2,3) is a multiplet of three real scalar field with a
constraint 
\begin{equation}
\Omega_{1}=n^{a}n^{a} - 1\approx 0.  \label{c1}
\end{equation}
Here, one notes that the unit 3-vector $n^a$ parametrizes an internal space $%
S^2$. One can now obtain the canonical Hamiltonian by performing the
Legendre transformation, 
\begin{equation}
H_c=\int {\rm d}^{2}x\left[\frac{f}{2}\pi^{a}\pi^{a} +\frac{1}{2f}%
(\partial_{i}n^{a})(\partial_{i}n^{a})\right]  \label{hc}
\end{equation}
where $\pi^{a}$ is the canonical momenta conjugate to the real scalar fields 
$n^{a}$ given by 
\begin{equation}
\pi^{a}=\frac{1}{f}\dot{n}^{a}.
\end{equation}
The time evolution of the constraint $\Omega_1$ yields one additional secondary 
constraint as follows, 
\begin{equation}
\Omega_{2}=n^{a}\pi^{a}\approx 0  \label{const22}
\end{equation}
and they form the second class constraints algebra 
\begin{equation}
\Delta_{kk^{\prime}}(x,y)=\{\Omega_{k}(x),\Omega_{k^{\prime}}(y)\}
=2\epsilon^{kk^{\prime}}n^{a}n^{a}\delta(x-y)  \label{delta}
\end{equation}
with $\epsilon^{12}=-\epsilon^{21}=1$.

Following the BFT formalism \cite{BFT,BFT1,kpr} which systematically
converts the second class constraints into first class ones, we introduce
two auxiliary fields $\Phi^{i}$ according to the number of second class
constraints $\Omega_{i}$ with the Poisson brackets 
\begin{equation}
\{\Phi^{i}(x), \Phi^{j}(y)\}=\omega^{ij}(x,y)  \label{phii}
\end{equation}
where we are free to make a choice 
\begin{equation}
\omega^{ij}(x,y)=\epsilon^{ij}\delta(x-y).  \label{2ep}
\end{equation}
The first class constraints $\tilde{\Omega}_{i}$ are then constructed as a
power series of the auxiliary fields: 
\begin{equation}
\tilde{\Omega}_{i}=\sum_{n=0}^{\infty}\Omega_{i}^{(n)},~~~~
\Omega_{i}^{(0)}=\Omega_{i}  \label{tilin}
\end{equation}
where $\Omega_{i}^{(n)}$ are polynomials in the auxiliary fields $\Phi^{j}$
of degree $n$, to be determined by the requirement that the first class
constraints $\tilde{\Omega}_{i}$ satisfy the closed algebra as follows 
\begin{equation}
\{\tilde{\Omega}_{i}(x),\tilde{\Omega}_{j}(y)\}=0.  \label{cijk}
\end{equation}
Since $\Omega_{i}^{(1)}$ are linear in the auxiliary fields, one can make
the ansatz 
\begin{equation}
\Omega_{i}^{(1)}(x)=\int {\rm d}^{2}y X_{ij}(x,y)\Phi^{j}(y).  \label{xijphi}
\end{equation}
Substituting Eq. (\ref{xijphi}) into Eq. (\ref{cijk}) leads to the following
relation 
\begin{equation}
\Delta_{ij}(x,y)+\int {\rm d}^{2}z {\rm d}^{2}z^{\prime}X_{ik}(x,z)
\omega^{kl}(z,z^{\prime})X_{jl}(z^{\prime},y)=0.  \label{delx}
\end{equation}
which, for the choice of Eq. (\ref{2ep}), has a solution 
\begin{equation}
X_{ij}(x,y)=\left( 
\begin{array}{cc}
2 & 0 \\ 
0 & -n^{a}n^{a}
\end{array}
\right)\delta (x-y).  \label{xij}
\end{equation}
Substituting Eq. (\ref{xij}) into Eqs. (\ref{tilin}) and (\ref{xijphi}) and
iterating this procedure, one obtains the first class constraints as follows
\begin{eqnarray}
\tilde{\Omega}_{1}&=&\Omega_{1}+2\Phi^{1},  \nonumber \\
\tilde{\Omega}_{2}&=&\Omega_{2}-n^{a}n^{a}\Phi^{2}  \label{1stconst}
\end{eqnarray}
which yield the strongly involutive first class constraint algebra (\ref
{cijk}). Therefore, we formally converted the second class constraint system
into the first class one. In the next section, we shall obtain the first
class physical fields and Hamiltonian.


\section{First class physical fields and Hamiltonian}
\setcounter{equation}{0}
\renewcommand{\theequation}{\arabic{section}.\arabic{equation}}


Now, following the improved BFT formalism\cite{kpr}, we newly construct the 
first class BFT physical fields $\tilde{{\cal F}}=(\tilde{n}^{a}, 
\tilde{\pi}^{a})$ corresponding to the original fields
defined by ${\cal F}=(n^{a},\pi^{a})$ in the extended phase space, which are
obtained as a power series in the auxiliary fields $\Phi^{i}$ by demanding
that they are strongly involutive: $\{\tilde{\Omega}_{i}, \tilde{{\cal F}}%
\}=0$. In general, the first class fields satisfying the boundary conditions 
$\tilde{{\cal F}}[{\cal F};0]={\cal F}$ can be found as 
\begin{equation}
\tilde{{\cal F}}[{\cal F};\Phi]={\cal F}+\sum_{n=1}^{\infty} \tilde{{\cal F}}%
^{(n)},~~~ \tilde{{\cal F}}^{(n)}\sim (\Phi)^{n}
\end{equation}
where the $(n+1)$-th order of iteration terms are given by the formula 
\begin{equation}
\tilde{{\cal F}}^{(n+1)}=-\frac{1}{n+1}\int {\rm d}^{2}x {\rm d}^{2}y {\rm d}%
^{2} z \Phi^{i}(x)\omega_{ij}(x,y)X^{jk}(y,z)G_{k}^{(n)}(z)
\end{equation}
with 
\begin{equation}
G_{i}^{(n)}(x)=\sum_{m=0}^{n}\{\Omega_{i}^{(n-m)},\tilde{{\cal F}}^{(m)}
\}_{({\cal F)}}+\sum_{m=0}^{n-2}\{\Omega_{i}^{(n-m)},\tilde{{\cal F}}%
^{(m+2)} \}_{(\Phi)}+\{\Omega_{i}^{(n+1)},\tilde{{\cal F}}^{(1)}\}_{(\Phi)}.
\end{equation}
After some lengthy algebra, we obtain the first class physical fields as 
\begin{eqnarray}
\tilde{n}^{a}&=&n^{a}\left(1-\sum_{n=1}^{\infty}\frac{(-1)^{n}(2n-3)!!}{n!} 
\frac{(\Phi^{1})^{n}}{(n^{a}n^{a})^{n}}\right),  \nonumber \\
\tilde{\pi}^{a}&=&\left(\pi^{a}-n^{a}\Phi^{2}\right)\left(1+\sum_{n=1}^{%
\infty} \frac{(-1)^{n}(2n-1)!!}{n!} \frac{(\Phi^{1})^{n}}{(n^{a}n^{a})^{n}}%
\right)  \label{pitilde}
\end{eqnarray}
with $(-1)!!=1$. Here one notes that the product of two polynomials in $%
\tilde{n}^{a}$ and $\tilde{\pi}^{a}$ yields unity, i.e., 
\begin{equation}
\left(1-\sum_{n=1}^{\infty}\frac{(-1)^{n}(2n-3)!!}{n!}\frac{(\Phi^{1})^{n}} {%
(n^{a}n^{a})^{n}}\right)\left(1+\sum_{n=1}^{\infty}\frac{(-1)^{n}(2n-1)!!}{n!%
} \frac{(\Phi^{1})^{n}}{(n^{a}n^{a})^{n}}\right)=1,
\end{equation}
which is crucial in derivation of the forthcoming equations such as Eqs. (%
\ref{commst}) and (\ref{oott}).

Using the novel property\cite{kpr} that any functional ${\cal K}(\tilde{%
{\cal F}})$ of the first class fields $\tilde{{\cal F}}$ is also first class, 
i.e., 
\begin{equation}
\tilde{{\cal K}}({\cal F};\Phi )={\cal K}(\tilde{{\cal F}})  \label{ktilde}
\end{equation}
we construct the first class Hamiltonian in terms of the above BFT physical
variables as follows 
\begin{equation}
\tilde{H}=\int {\rm d}^{2}x\left[\frac{f}{2}\tilde{\pi}^{a}\tilde{\pi}^{a} +%
\frac{1}{2f}(\partial_{i}\tilde{n}^{a})(\partial_{i}\tilde{n}^{a})\right].
\label{htilde}
\end{equation}
We then directly rewrite this Hamiltonian in terms of original fields
and auxiliary ones\footnote{%
In deriving the first class Hamiltonian $\tilde{H}$ of Eq. (\ref{hct}), we
have used the conformal map condition, $n^{a}\partial_{i} n^{a}=0$, which
states that the radial vector is perpendicular to the tangent on the $S^{2}$
sphere in the extended phase space of the O(3) nonlinear sigma model. The
geometrical structure is then conserved in the map from the original phase
space to the extended one. In other words, the $S^{2}$ sphere given by $%
n^{a}n^{a}=1$ in the original phase space is casted into the other sphere $%
n^{a}n^{a}=1-2\Phi^{1}$ in the extended phase space without any distortion.
With this conformal map condition, we have confirmed that the first class
Hamiltonian $\tilde{H}$ of Eq. (\ref{hct}) is consistently derived from both
the novel property (\ref{ktilde}) and the infinitely iterated standard
procedure, as expected.} 
\begin{eqnarray}
\tilde{H}&=&\int {\rm d}^{2}x~\left[\frac{f}{2}(\pi^{a}-n^{a}\Phi^{2})
(\pi^{a}-n^{a}\Phi^{2})\frac{n^{c}n^{c}}{n^{c}n^{c}+2\Phi^{1}} \right. 
\nonumber \\
& &\left.+\frac{1}{2f}(\partial_{i}n^{a})(\partial_{i}n^{a})\frac{%
n^{c}n^{c}+2 \Phi^{1}}{n^{c}n^{c}}\right],  \label{hct}
\end{eqnarray}
which is strongly involutive with the first class constraints 
\[
\{\tilde{\Omega}_{i},\tilde{H}\}=0. 
\]

It seems to be appropriate to comment on the Hamiltonian (\ref{hct}). The
form of the first term in this Hamiltonian is exactly the same as that of
the SU(2) Skyrmion\cite{sk2,su2bft}. 

On the other hand, with the first class Hamiltonian (\ref{hct}), one cannot 
naturally generate the first class Gauss' law constraint from the time 
evolution of the constraint $\tilde{\Omega}_{1}$.  Now, by introducing an 
additional term proportional to the first class constraints 
$\tilde{\Omega}_{2}$ into $\tilde{H}$, we obtain an equivalent first class 
Hamiltonian 
\begin{equation}
\tilde{H}^{\prime}=\tilde{H}+\int {\rm d}^{2}x f\Phi^{2}\tilde{\Omega}_{2}  
\label{hctp}
\end{equation}
which naturally generates the Gauss' law constraint 
\begin{eqnarray}
\{\tilde{\Omega}_{1},\tilde{H}^{\prime}\}&=&2f\tilde{\Omega}_{2},  \nonumber
\\
\{\tilde{\Omega}_{2},\tilde{H}^{\prime}\}&=&0.
\end{eqnarray}
Here one notes that $\tilde{H}$ and $\tilde{H}^{\prime}$ act on physical
states in the same way since such states are annihilated by the first class
constraints. Similarly, the equations of motion for observables will also be
unaffected by this difference. Furthermore, if we take the limit $%
\Phi^{i}\rightarrow 0$, then our first class system exactly returns to the
original second class one.


\section{Structures of Dirac and Poisson brackets}
\setcounter{equation}{0}
\renewcommand{\theequation}{\arabic{section}.\arabic{equation}}


Next let us consider the Poisson brackets of fields in the extended phase
space $\tilde{{\cal F}}$ and identify the Dirac brackets by taking the
vanishing limit of auxiliary fields. After some manipulation from Eq. (\ref
{pitilde}), one could obtain the commutators 
\begin{eqnarray}
\{\tilde{n}^{a}(x),\tilde{n}^{b}(y)\}&=&0,  \nonumber \\
\{\tilde{n}^{a}(x),\tilde{\pi}^{b}(y)\}&=&(\delta^{ab}-\frac{\tilde{n}^{a} 
\tilde{n}^{b}}{\tilde{n}^{c}\tilde{n}^{c}})\delta(x-y),  \nonumber \\
\{\tilde{\pi}^{a}(x),\tilde{\pi}^{b}(y)\}&=&\frac{1}{\tilde{n}^{c} \tilde{n}%
^{c}}(\tilde{n}^{b}\tilde{\pi}^{a} -\tilde{n}^{a}\tilde{\pi}^{b})\delta
(x-y).  \label{commst}
\end{eqnarray}
In the limit $\Phi^{i}\rightarrow 0$, the above Poisson brackets in the
extended phase space exactly reproduce the corresponding Dirac brackets \cite
{bo} 
\begin{eqnarray}
\{\tilde{n}^{a},\tilde{n}^{b}\}|_{\Phi =0}&=& \{n^{a},n^{b}\}_{D},  \nonumber
\\
\{\tilde{n}^{a},\tilde{\pi}^{b}\}|_{\Phi =0}&=& \{n^{a},\pi^{b}\}_{D}, 
\nonumber \\
\{\tilde{\pi}^{a},\tilde{\pi}^{b}\}|_{\Phi =0}&=& \{\pi^{a},\pi^{b}\}_{D}
\label{commstd}
\end{eqnarray}
where 
\begin{equation}
\{A(x),B(y)\}_{D}=\{A(x),B(y)\}-\int d^2z d^2 z^{\prime}
\{A(x),\Omega_{k}(z)\}\Delta^{k k^{\prime}}
\{\Omega_{k^{\prime}}(z^{\prime}),B(y)\}
\end{equation}
with $\Delta^{k k^{\prime}}$ being the inverse of $\Delta_{k k^{\prime}}$ in
Eq. (\ref{delta}). Also it is amusing to see in Eq. (\ref{commst}) that
these Poisson brackets of $\tilde{{\cal F}}$'s have exactly the same form of
the Dirac brackets of the field ${\cal F}$ obtained by the replacement of $%
{\cal F}$ with $\tilde{{\cal F}}$. In other words, the functional $\tilde{%
{\cal K}}$ in Eq. (\ref{ktilde}) corresponds to the Dirac brackets $%
\{A,B\}|_{D}$ and hence $\tilde{{\cal K}}$ corresponding to $\{\tilde{A},%
\tilde{B}\}$ becomes 
\begin{equation}
\{\tilde{A},\tilde{B}\}=\{A,B\}_{D}|_{A\rightarrow \tilde{A},B\rightarrow 
\tilde{B}}.
\end{equation}
This kind of situation happens again when one considers the first class
constraints (\ref{1stconst}). More precisely these first class constraints
in the extended phase space can be rewritten as 
\begin{eqnarray}
\tilde{\Omega}_{1}&=&\tilde{n}^{a}\tilde{n}^{a}-1,  \nonumber \\
\tilde{\Omega}_{2}&=&\tilde{n}^{a}\tilde{\pi}^{a},  \label{oott}
\end{eqnarray}
which are form-invariant with respect to the second class constraints (\ref
{const22}).


\section{BFV-BRST gauge fixing}
\setcounter{equation}{0}
\renewcommand{\theequation}{\arabic{section}.\arabic{equation}}


In this section, in order to obtain the effective Lagrangian, we introduce
two canonical sets of ghosts and anti-ghosts together with auxiliary fields
in the framework of the BFV formalism \cite{bfv,fik,biz}, which is
applicable to theories with the first class constraints: 
\[
({\cal C}^{i},\bar{{\cal P}}_{i}),~~({\cal P}^{i}, \bar{{\cal C}}_{i}),
~~(N^{i},B_{i}),~~~~(i=1,2) 
\]
which satisfy the super-Poisson algebra 
\[
\{{\cal C}^{i}(x),\bar{{\cal P}}_{j}(y)\}=\{{\cal P}^{i}(x), \bar{{\cal C}}%
_{j}(y)\}=\{N^{i}(x),B_{j}(y)\}=\delta_{j}^{i}\delta(x-y), 
\]
where the super-Poisson bracket is defined as 
\[
\{A,B\}=\frac{\delta A}{\delta q}|_{r}\frac{\delta B}{\delta p}|_{l}
-(-1)^{\eta_{A}\eta_{B}}\frac{\delta B}{\delta q}|_{r}\frac{\delta A} {%
\delta p}|_{l} 
\]
where $\eta_{A}$ denotes the number of fermions called ghost number in $A$
and the subscript $r$ and $l$ imply right and left derivatives, respectively.

In the O(3) nonlinear sigma model, the nilpotent BRST charge $Q$, the
fermionic gauge fixing function $\Psi$ and the BRST invariant minimal
Hamiltonian $H_{m}$ are given by 
\begin{eqnarray}
Q&=&\int {\rm d}^{2}x~({\cal C}^{i}\tilde{\Omega}_{i}+{\cal P}^{i}B_{i}), 
\nonumber \\
\Psi&=&\int {\rm d}^{2}x~(\bar{{\cal C}}_{i}\chi^{i}+\bar{{\cal P}}%
_{i}N^{i}),  \nonumber \\
H_{m}&=&\tilde{H}^{\prime}-\int {\rm d}^{2}x~2f{\cal C}^{1}\bar{{\cal P}}%
_{2},
\end{eqnarray}
which satisfy the following relations 
\begin{equation}
\{Q,H_{m}\}=0,~~Q^{2}=\{Q,Q\}=0,~~\{\{\Psi,Q\},Q\}=0.
\end{equation}
The effective quantum Lagrangian is then described as 
\begin{equation}
L_{eff}=\int {\rm d}^{2}x~(\pi^{a}\dot{n}^{a}+\pi_{\theta}\dot{\theta} +B_{2}%
\dot{N}^{2}+\bar{{\cal P}}_{i}\dot{{\cal C}}^{i}+\bar{{\cal C}}_{2} \dot{%
{\cal P}}^{2})-H_{tot}
\end{equation}
with $H_{tot}=H_{m}-\{Q,\Psi\}$. Here we identified the auxiliary fields $%
\Phi^{i}$ with a canonical conjugate pair $(\theta,\pi_{\theta})$, i.e., 
\begin{equation}
\Phi^{i}=(\theta,\pi_{\theta})
\end{equation}
and $\int {\rm d}^{2}x~(B_{1}\dot{N}^{1} +\bar{{\cal C}}_{1}\dot{{\cal P}}%
^{1})=\{Q,\int{\rm d}^{2}x~\bar{{\cal C}}_{1} \dot{N}^{1}\}$ terms are
suppressed by replacing $\chi^{1}$ with $\chi^{1} +\dot{N}^{1}$.

Now we choose the unitary gauge 
\begin{equation}
\chi^{1}=\Omega_{1},~~~\chi^{2}=\Omega_{2}
\end{equation}
and perform the path integration over the fields $B_{1}$, $N^{1}$, $\bar{%
{\cal C}}_{1}$, ${\cal P}^{1}$, $\bar{{\cal P}}_{1}$ and ${\cal C}^{1}$, by
using the equations of motion, to yield the effective Lagrangian of the form 
\begin{eqnarray}
L_{eff}&=&\int{\rm d}^{2}x~\left[\pi^{a}\dot{n}^{a}+\pi_{\theta}\dot{\theta}
+B\dot{N}+\bar{{\cal P}}\dot{{\cal C}}+\bar{{\cal C}}\dot{{\cal P}}\right. 
\nonumber \\
& &\left.-\frac{f}{2}(\pi^{a}-n^{a}\pi_{\theta})(\pi^{a}-n^{a}\pi_{\theta}) 
\frac{n^{c}n^{c}}{n^{c}n^{c}+2\theta}-\frac{1}{2f}(\partial_{i}n^{a})
(\partial_{i}n^{a})\frac{n^{c}n^{c}+2\theta}{n^{c}n^{c}}\right.  \nonumber \\
& &\left.-f\pi_{\theta}\tilde{\Omega}_{2}+2n^{a}n^{a}\pi_{\theta}\bar{{\cal C%
}} {\cal C}+\tilde{\Omega}_{2}N+B\Omega_{2}+\bar{{\cal P}}{\cal P}\right]
\end{eqnarray}
with redefinitions: $N\equiv N^{2}$, $B\equiv B_{2}$, $\bar{{\cal C}}\equiv 
\bar{{\cal C}}_{2}$, ${\cal C}\equiv {\cal C}^{2}$, $\bar{{\cal P}}\equiv 
\bar{{\cal P}}_{2}$, ${\cal P}\equiv {\cal P}_{2}$.

Using the variations with respect to $\pi^{a}$, $\pi_{\theta}$, ${\cal P}$
and $\bar{{\cal P}}$, one obtain the relations 
\begin{eqnarray}
\dot{n}^{a}&=&f(\pi^{a}-n^{a}\pi_{\theta})n^{c}n^{c}+n^{a}(f\pi_{%
\theta}-N-B),  \nonumber \\
\dot{\theta}&=&-f n^{a}(\pi^{a}-n^{a}\pi_{\theta})n^{c}n^{c}+n^{a}n^{a}
(-2f\pi_{\theta}-2\bar{{\cal C}}{\cal C}+N)+f n^{a}\pi^{a},  \nonumber \\
{\cal P}&=&-\dot{{\cal C}},~~~~~\bar{{\cal P}}=\dot{\bar{{\cal C}}}
\end{eqnarray}
to yield the effective Lagrangian 
\begin{eqnarray}
L_{eff}&=&\int{\rm d}^{2}x~\left[\frac{1}{2f n^{c}n^{c}}(\partial_{%
\mu}n^{a}) (\partial^{\mu}n^{a})-\frac{1}{2f}\left(\frac{\dot{\theta}}{%
n^{c}n^{c}} +(B+2\bar{{\cal C}}{\cal C})n^{c}n^{c}\right)^{2}\right. 
\nonumber \\
& &\left. +\frac{1}{fn^{c}n^{c}}n^{a}\{\dot{n}^{a}+n^{a}(\frac{\dot{\theta}%
} {n^{c}n^{c}}+(B+2\bar{{\cal C}}{\cal C})n^{c}n^{c})\}(B+N) \right. 
\nonumber \\
& &\left. +B\dot{N}+\dot{\bar{{\cal C}}}\dot{{\cal C}}\right].
\end{eqnarray}

After identifying 
\begin{equation}
N=-B+\frac{\dot{\theta}}{n^{c}n^{c}},
\end{equation}
we then obtained the effective Lagrangian of the form 
\begin{equation}
L_{eff}= L_0 + L_{WZ} + L_{gh}
\end{equation}
where 
\begin{eqnarray}
L_{WZ}&=&\int{\rm d}^{2}x~\left[\frac{1}{fn^{c}n^{c}}(\partial_{\mu}n^{a})
(\partial^{\mu}n^{a}){\theta}-\frac{1}{2f(n^{c}n^{c})^{2}}\dot{\theta}^{2} %
\right], \\
L_{gh}&=&\int{\rm d}^{2}x~\left[-\frac{1}{2f}(n^{a}n^{a})^{2}(B+2\bar{{\cal C%
}}{\cal C})^{2} -\frac{\dot{\theta}\dot{B}}{n^{c}n^{c}} +\dot{\bar{{\cal C}}}%
\dot{{\cal C}}\right].
\end{eqnarray}
This Lagrangian is invariant under the BRST transformation 
\begin{eqnarray}
\delta_{B}n^{a}&=&\lambda n^{a}{\cal C},~~~ \delta_{B}\theta=-\lambda
n^{a}n^{a}{\cal C},  \nonumber \\
\delta_{B}\bar{{\cal C}}&=&-\lambda B,~~~ \delta_{B}{\cal C}=\delta_{B}B=0.
\end{eqnarray}
This completes the standard procedure of BRST invariant gauge fixing in the
BFV formalism.

It seems to be appropriate to comment that we could directly read off the
gauge-invariant first class Lagrangian corresponding to the first class
Hamiltonian (23) from Eq. (39) as follows 
\begin{equation}
\tilde{L}^{\prime}= L_0 + L_{WZ}  \label{1stlag}
\end{equation}


\section{Semi-classical quantization}
\setcounter{equation}{0}
\renewcommand{\theequation}{\arabic{section}.\arabic{equation}}


In this section, we perform a semi-classical quantization of the $Q=1$
sector of the O(3) nonlinear sigma model \cite{bo} to consider physical
aspects of the theory. Here the topological charge $Q$ is given by 
\begin{equation}
Q=\int {\rm d}^{2}x J_{0}(x),
\end{equation}
where $J^{\mu}$ is the topological current 
\begin{equation}
J_{\mu}=\frac{1}{8\pi}\epsilon_{\mu\nu\lambda}\epsilon^{abc}n^{a}\partial^{%
\nu} n^{b}\partial^{\lambda}n^{c}.
\end{equation}

On the other hand, the O(3) nonlinear soliton theory is not invariant under
spatial rotations $U_{J}(1)$ or isospin rotation $U_{I}(1)$ separately. It
is instead invariant under a combined spatial and isospin rotation. Since
the theory is invariant under $U_{I}(1)\times U_{J}(1)$, as a first
approximation to the quantum ground state we could quantize zero modes
responsible for classical degeneracy by introducing collective coordinates
as follows 
\footnote{%
Given the soliton configuration (\ref{conf}), one could easily calculate the
spatial derivatives to yield $n^{a}\partial_{i}n^{a}=0$. As a result one
could conclude that the conformal map condition is fulfilled in the above
soliton configuration, as expected.
On the other hand we also have the relation $n^{a}\partial_{0}n^{a}=0$.
}
\begin{eqnarray}
n^{1}&=&\cos (\alpha (t)+\phi )\sin F(r),  \nonumber \\
n^{2}&=&\sin (\alpha (t)+\phi )\sin F(r),  \nonumber \\
n^{3}&=&\cos F(r),  \label{conf}
\end{eqnarray}
where $(r,\phi)$ are the polar coordinates and $\alpha (t)$ is the
collective coordinates. Here, in order to ensure $Q=1$, the profile function 
$F(r)$ satisfies the boundary conditions: $\lim_{r\rightarrow \infty}F(r)=\pi
$ and $F(0)=0$.

Using the above soliton configuration, we obtain the unconstrained
Lagrangian of the form 
\begin{equation}
L=-E+\frac{1}{2}{\cal I}\dot{\alpha}^{2}  \label{originl}
\end{equation}
where the soliton energy and the moment of inertia are given by 
\begin{eqnarray}
E&=&\frac{\pi}{f}\int_{0}^{\infty}{\rm d}r r \left[(\frac{{\rm d}F} {{\rm d}r%
})^{2}+\frac{\sin^{2}F}{r^{2}}\right],  \nonumber \\
{\cal I}&=&\frac{2\pi}{f}\int_{0}^{\infty}{\rm d}r r \sin^{2}F.
\end{eqnarray}
Introducing the canonical momentum conjugate to the collective coordinate $%
\alpha$ 
\begin{equation}
p_{\alpha}={\cal I}\dot{\alpha}
\end{equation}
we then have the canonical Hamiltonian 
\begin{equation}
H=E+\frac{1}{2{\cal I}}p_{\alpha}^{2}.  \label{hcc}
\end{equation}

At this stage, one can associate the Hamiltonian (\ref{hcc}) with the
previous one (\ref{hc}), which was given by the canonical momenta $\pi^{a}$.
Given the soliton configuration (\ref{conf}) one can obtain the relation
between $\pi^{a}$ and $p_{\alpha}$ as follows 
\begin{equation}
\pi^{a}\pi^{a}=\frac{\sin^{2}F}{{\cal I}^{2}f^{2}}p_{\alpha}^{2}  \label{rel}
\end{equation}
to yield the integral 
\begin{equation}
\int {\rm d}^{2}x \frac{f}{2}\pi^{a}\pi^{a}=\frac{1}{2{\cal I}}%
p_{\alpha}^{2}.  \label{integ}
\end{equation}
Since the spatial derivative term in (\ref{hc}) yields nothing but the
soliton energy $E$, one can easily see, together with the relation (\ref
{integ}), that the canonical Hamiltonian (\ref{hc}) is equivalent to the
other one (\ref{hcc}), as expected.

Now, let us define the angular momentum operator $J$ as follows 
\begin{equation}
J=\int {\rm d}^{2}x \epsilon_{ij}x^{i}T^{oj}  \label{jj}
\end{equation}
where the symmetric energy-momentum tensor is given by 
\begin{eqnarray}
T^{\mu\nu}&=&\frac{\partial{\cal L}}{\partial(\partial_{\mu}n^{a})}
\partial^{\nu}n^{a}-g^{\mu\nu}{\cal L}  \nonumber \\
&=&\frac{1}{f}\partial^{\mu}n^{a}\partial^{\nu}n^{a}-\frac{1}{2f}g^{\mu\nu}
\partial_{\sigma}n^{a}\partial^{\sigma}n^{a}.  \label{tt}
\end{eqnarray}

Then, substituting the configuration (\ref{conf}) into Eq. (\ref{tt}), we
obtain the angular momentum operator of the form 
\begin{equation}
J=-{\cal I}\dot{\alpha}=-p_{\alpha}.
\end{equation}
With $p_{\alpha}=-i\frac{\partial}{\partial \alpha}$, the angular momentum
operator can be rewritten in terms of the U(1) isospin operator $I$ as
follows 
\begin{equation}
J=i\frac{\partial}{\partial \alpha}=I
\end{equation}
to yield the Hamiltonian of the form 
\begin{equation}
H=E+\frac{1}{2{\cal I}}J^{2}.  \label{jham}
\end{equation}
Here one notes that the above Hamiltonian can be interpreted as that of a
rigid rotator.

Next let us consider the zero modes in the extended phase space by introducing
the soliton configuration
\begin{eqnarray}
n^{1}&=&(1-2\theta)^{1/2}\cos (\alpha (t)+\phi )\sin F(r),  \nonumber \\
n^{2}&=&(1-2\theta)^{1/2}\sin (\alpha (t)+\phi )\sin F(r),  \nonumber \\
n^{3}&=&(1-2\theta)^{1/2}\cos F(r),  \label{conf2}
\end{eqnarray}
which satisfy the first class constraint $n^{a}n^{a}=1-2\theta$ of Eq. 
(\ref{1stconst}).\footnote{Due to the identity 
$\left(1-\sum_{n=1}^{\infty}\frac{(-1)^{n}(2n-3)!!}{n!}\frac{\theta ^{n}}
{(n^{a}n^{a})^{n}}\right)^{2}=\frac{1}{1-2\theta}$, one can
easily see that the first class physical fields $\tilde{n}^{a}$ of Eq. 
(\ref{pitilde}) satisfy the corresponding first class constraint 
$\tilde{n}^{a}\tilde{n}^{a}=1$ of Eq. (\ref{oott}).}  In this configuration 
from Eqs. (2.1) and (5.11) we then obtain 
\begin{eqnarray}
L_{0}&=& -({1-2\theta})E + \frac{1}{2} {\cal I} ({1-2\theta})\dot{\alpha}^{2} 
+ \int{\rm d}^{2}x\frac{1}{2f}\frac{\dot{\theta}^{2}}{1-2\theta},  \nonumber \\
L_{WZ}&=& -{2\theta}E + \frac{1}{2} {\cal I} ({2\theta})\dot{\alpha}^{2} 
- \int{\rm d}^{2}x\frac{1}{2f}\frac{\dot{\theta}^{2}}{1-2\theta}.
\end{eqnarray}
to yield the first class Lagrangian $\tilde{L}^{\prime}$ in (\ref{1stlag}), 
which is remarkably the Lagrangian (\ref{originl}) given in the original phase 
space due to the exact cancellation of $\theta$-terms.  Consequently the 
quantization of zero modes in the extended phase space reproduces the same 
energy spectrum (\ref{jham}). This phenomenon originates from the fact that 
the collective coordinates $\alpha$ in the Lagrangian (\ref{originl}) are not 
affected by the constraints (\ref{c1}) and (\ref{1stconst}) for the real 
scalar fields $n^{a}$. Here one notes that in the SU(2) Skyrmion model the 
collective coordinates themselves are constrained to yield the modified energy 
spectrum\cite{sk2,sk} in contrast to the case of the O(3) nonlinear sigma 
model.


\section{Conclusion}


In summary, we have constructed the first class BFT physical fields, in
terms of which the first class Hamiltonian is formulated to be consistent
with the Hamiltonian with the original fields and auxiliary fields. The
Poisson brackets of the BFT physical fields are also built to reproduce the
corresponding Dirac brackets in the limit of vanishing auxiliary fields. In
the Batalin, Fradkin and Vilkovisky (BFV) scheme\cite{bfv,fik,biz}, we have
then obtained the BRST invariant gauge fixed Lagrangian including the
(anti)ghost fields, and its BRST transformation rules. On the other hand,
introducing the collective coordinates in the soliton configuration, we have
performed the semi-classical quantization to yield the energy spectrum which
can be interpreted as that of the rigid rotator. Also, we have newly 
recognized that the spectrum of zero modes are reproduced through the BFT 
Hamiltonian procedure. The Chern-Simons terms will be included through further 
investigation to generalize the BFT and BFV-BRST schemes for the nonlinear 
sigma model to the fractional spin systems.

\vskip 1.0cm 
One of us (S.T.H.) would like to thank G.E. Brown at Stony Brook for constant 
concerns and encouragement. We would also like to thank Y.-W. Kim and M.-I. 
Park for valuable discussions. The present work was supported by the Basic 
Science Research Institute Program, Korean Research Foundation, Project No. 
1998-015-D00074.


\begin{thebibliography}{99}
\bibitem{eucl}  C. Domb and M.S. Green, Phase Transitions and Critical
Phenomena (Academic, New York, 1972).

\bibitem{min}  F.D.M. Haldane, Phys. Lett. A93 (1983) 464; Phys. Rev. Lett.
50 (1983) 1153.

\bibitem{cs}  F. Wilczek and A. Zee, Phys. Rev. Lett. 51 (1983) 2250; A.M.
Polyakov, Mod. Phys. Lett. A3 (1999) 417; Y. Wu and A. Zee, Phys. Lett. B147
(1984) 325; G. Semenoff, Phys. Rev. Lett. 61 (1988) 517.

\bibitem{bo}  M. Bowick, D. Karabali, L.C.R. Wijewardhana, Nucl. Phys. B271
(1986) 417.

\bibitem{di}  P.A.M. Dirac, Lectures in Quantum Mechanics (Yeshiva
University, New York, 1964).

\bibitem{brst}  C. Becci, A. Rouet, R. Stora: Ann. Phys. [NY] {\bf 98}
(1976) 287; I.V. Tyutin: Lebedev Preprint 39 (1975)

\bibitem{BFT}  I.A. Batalin, E.S. Fradkin, Phys. Lett. B180 (1986) 157;
Nucl. Phys. B279 (1987) 514; I.A. Batalin, I.V. Tyutin, Int. J. Mod. Phys.
A6 (1991) 3255.

\bibitem{BFT1}  R. Banerjee, Phys. Rev. D48 (1993) R5467; W.T. Kim, Y.-J.
Park, Phys. Lett. B336 (1994) 376.

\bibitem{kpr}  Y.-W. Kim, Y.-J. Park, K.D. Rothe, J. Phys. G24 (1998) 953;
Y.-W. Kim, K. D. Rothe, Nucl. Phys. B510 (1998) 511; M.-I. Park, Y.-J. Park,
Int. J. Mod. Phys. A13 (1998) 2179.

\bibitem{sk2}  S.-T. Hong, Y.-W. Kim, Y.-J. Park, Sogang Univ. Preprint
SOGANG-HEP 249/98, {\tt hep-th/9811066} (1998) to appear in Phys. Rev. D.

\bibitem{su2bft}  S.-T. Hong, Y.-W. Kim, Y.-J. Park, Sogang Univ. Preprint
SOGANG-HEP 256/99 (1999).

\bibitem{bfv}  E.S. Fradkin, G.A. Vilkovisky, Phys. Lett. B55 (1975) 224; M.
Henneaux, Phys. Rep. C126 (1985) 1.

\bibitem{fik}  T. Fujiwara, Y. Igarashi, J. Kubo, Nucl. Phys. B341 (1990)
695; Y.-W. Kim, S.-K. Kim, W. T. Kim, Y.-J. Park, K. Y. Kim, Y. Kim, Phys.
Rev. D46 (1992) 4574.

\bibitem{biz}  C. Bizdadea, S.O. Saliu, Nucl. Phys. B456 (1995) 473.

\bibitem{bgb}  N. Banerjee, S. Ghosh, and R. Banerjee, Nucl. Phys. B417, 257
(1994).

\bibitem{sk}  W. Oliveira, J.A. Neto, Int. J. Mod. Phys. A12 (1997) 4895. 
\end{thebibliography}
\end{document}